\documentclass{article}

\usepackage{PRIMEarxiv}

\usepackage[utf8]{inputenc} 
\usepackage[T1]{fontenc}    
\usepackage{hyperref}       
\usepackage{url}            
\usepackage{booktabs}       
\usepackage{amsfonts}       
\usepackage{nicefrac}       
\usepackage{microtype}      
\usepackage{lipsum}
\usepackage{fancyhdr}       
\usepackage{graphicx}       
\graphicspath{{media/}}     

\usepackage{amsmath,amssymb,amsfonts}
\usepackage{algorithmic}
\usepackage{textcomp}
\usepackage{xcolor}
\usepackage{makecell}
\usepackage{multirow}

\usepackage{bm}
\usepackage{adjustbox}

\usepackage{caption}

\usepackage{tikz}

\newcommand{\tikzcmark}{%
\tikz[scale=0.23] {
    \draw[line width=0.7,line cap=round] (0.25,0) to [bend left=10] (1,1);
    \draw[line width=0.8,line cap=round] (0,0.35) to [bend right=1] (0.23,0);
}}

\title{GSBIQA: Green Saliency-guided Blind Image Quality Assessment Method}

\author{
  Zhanxuan Mei \\
  University of Southern California \\
  Los Angeles, USA\\
  \texttt{zhanxuan@usc.edu} \\
   \And
  Yun-Cheng Wang \\
  University of Southern California \\
  Los Angeles, USA\\
  \texttt{yunchenw@@usc.edu} \\
   \And
  C.-C. Jay Kuo \\
  University of Southern California \\
  Los Angeles, USA\\
  \texttt{cckuo@sipi.usc.edu} \\
}

\begin{document}
\maketitle

\begin{abstract}
Blind Image Quality Assessment (BIQA) is an essential task that estimates the perceptual quality of images without reference. 
While many BIQA methods employ deep neural networks (DNNs) and incorporate saliency detectors to enhance performance, their large model sizes limit deployment on resource-constrained devices.
To address this challenge, we introduce a novel and non-deep-learning BIQA method with a lightweight saliency detection module, called Green Saliency-guided Blind Image Quality Assessment (GSBIQA). 
It is characterized by its minimal model size, reduced computational demands, and robust performance. 
Experimental results show that the performance of
GSBIQA is comparable with state-of-the-art DL-based methods with significantly lower resource requirements.
\end{abstract}

\section{Introduction}\label{G_sec:introduction}

Objective image quality assessment (IQA) is pivotal in various multimedia applications. It can be categorized into three distinct types: Full-Reference IQA (FR-IQA), Reduced-Reference IQA (RR-IQA), and No-Reference IQA (NR-IQA). 
FR-IQA directly compares a distorted image against a reference or original image to assess quality. 
RR-IQA, on the other hand, uses partial information from the reference images to evaluate the quality of the target images. 
NR-IQA, also known as blind image quality assessment (BIQA), becomes essential in scenarios where reference images are unavailable, such as at the receiver's end or for user-generated content on social media. The demand for BIQA has surged with the increasing popularity of such platforms.

Research in BIQA has gained significant momentum over recent years, branching into two primary approaches: conventional methods and deep-learning-based (DL-based) methods. 
Conventional BIQA methods typically follow a structured pipeline involving quality-aware feature extraction followed by regression to map these features to quality scores. 
Over the past two decades, various conventional BIQA methods have been developed, including those based on Natural Scene Statistics (NSS)~\cite{mittal2012no} and codebook-based approaches~\cite{ye2012unsupervised}. More recently, inspired by the success of DNNs in computer vision, researchers have developed DL-based methods~\cite{kim2016fully, bosse2017deep} to solve the BIQA problem. 
However, the high cost of collecting large-scale annotated IQA datasets and the tendency of DL-based methods to overfit on limited-sized datasets pose significant challenges. 
Some advanced DL-based methods have adopted large pre-trained models from external datasets, such as ImageNet~\cite{deng2009imagenet} to address this. 
Based on that, a further direction~\cite{chetouani2018convolutional} is enhancing dataset variability by sampling patches randomly from quality-annotated images and incorporating saliency-guided approaches to improve accuracy, reflecting the intrinsic link between perceptual quality prediction and saliency prediction.

However, due to their large model sizes and high computational complexity, DL-based BIQA methods face substantial challenges, particularly in their deployment on mobile or edge devices. 
Moreover, the conventional saliency predictors often used in these methods do not significantly enhance perceptual quality predictions. 
In response to these challenges, this work introduces a novel, efficient, and saliency-assisted BIQA method named Green Saliency-guided Blind Image Quality Assessment (GSBIQA). It is particularly suited for real-world applications with authentic distortions and offers a modular design with a straightforward training pipeline.


We should point out that green image saliency detection (GreenSaliency) and green blind image quality assessment (GreenBIQA) were initially presented in \cite{mei2024lightweight} and \cite{mei2024greensaliency}, respectively. 
This work offers a significant expansion of these initial efforts by combining and extending these two lightweight models into a comprehensive and advanced BIQA method. 


\section{Related Work}\label{G_sec:related}

\subsection{Blind Image Quality Assessment Methods}
Over the past two decades, several conventional BIQA methods have been developed, including those based on Natural Scene Statistics (NSS)~\cite{mittal2012no} and codebook approaches~\cite{ye2012unsupervised}.
The remarkable success of DNNs has inspired a surge in research focused on employing DL-based methods to address the BIQA challenge. 
Early-stage DL-based BIQA methods, such as BIECON~\cite{kim2016fully} and WaDIQaM~\cite{bosse2017deep}, constructed neural networks consisting of convolutional layers, max pooling, and fully connected layers. 
To mitigate the issue of limited annotated datasets in IQA, most contemporary DL-based approaches leverage advanced pre-trained models for feature extraction. 
A probabilistic quality representation was introduced in PQR~\cite{zeng2018blind}, utilizing a robust and optimal loss function to more accurately reflect the score distribution provided by different evaluators, thereby enhancing prediction accuracy and expediting the training phase. 
HyperIQA~\cite{su2020blindly} employed a self-adaptive hypernetwork architecture that adjusts the quality prediction parameters. This architecture can manage a broad spectrum of distortions through a local distortion-aware module and adapts to diverse content through perceptual quality patterns. 
DBCNN~\cite{zhang2018blind} utilizes networks pre-trained on synthetic-distortion datasets and ImageNet for distortion and general classification, respectively. 
Features from both models are combined using bilinear pooling for final quality prediction. Addressing the large model size issue common in DL-based methods, GreenBIQA~\cite{mei2024lightweight} demonstrates a successful approach using a non-DL-based method to achieve high accuracy with a smaller model size and reduced computational complexity.

\begin{figure*}[!htbp]
\centering
\includegraphics[width=1\linewidth]{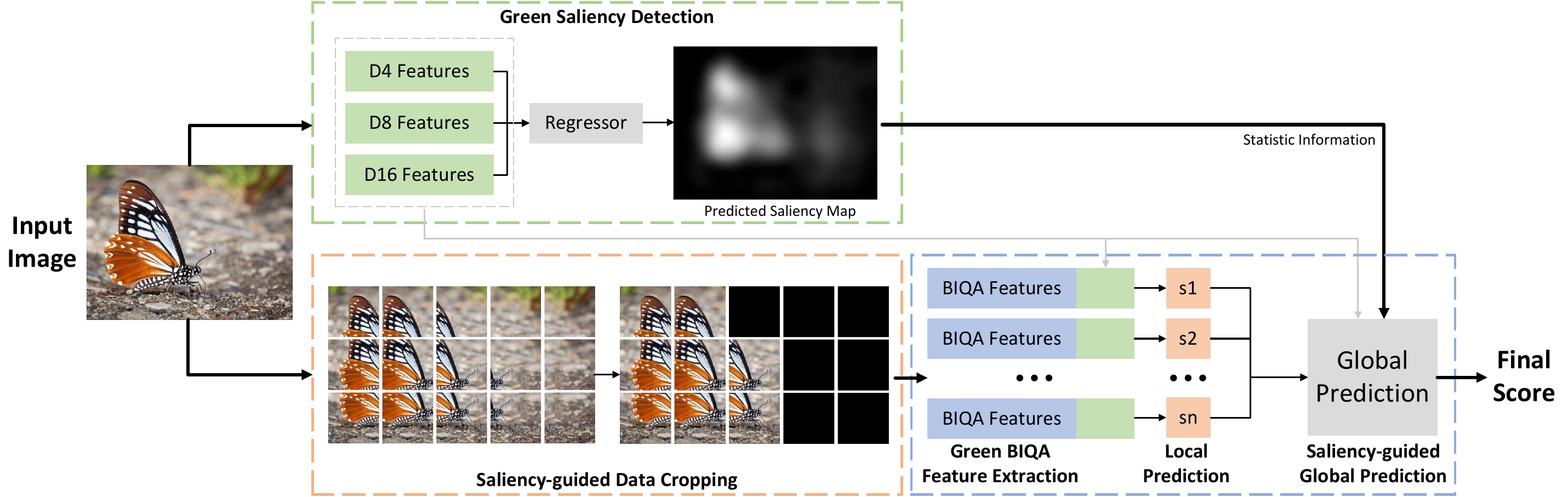}\\
\caption{An overview of the proposed GSBIQA method.}\label{fig:pipeline}
\end{figure*}

\subsection{Saliency-guided BIQA methods}

Image saliency detectors aim to elucidate the mechanisms underlying human observation, a goal that aligns closely with the objectives of IQA, particularly in the context of BIQA.
Consequently, some researchers have incorporated saliency detection techniques within BIQA frameworks to improve their efficacy. 
Within DL-based frameworks, saliency models enhance data pre-processing, such as in~\cite{chetouani2018convolutional}, where these models guide data cropping by identifying and selecting patches containing salient objects from the original images. 
Furthermore, saliency detectors are integrated into the data cropping and decision fusion processes, as demonstrated in~\cite{shen2022saliency}.
Rather than merely applying saliency models as either a pre-processing or post-processing tool, ~\cite{yang2019sgdnet} introduces SGDNet. 
This network is built on an end-to-end multi-task learning framework that addresses image saliency prediction and quality assessment. 
\section{Proposed Method}\label{G_sec:Proposed_method}
An overview of the proposed GSBIQA method is illustrated in Figure \ref{fig:pipeline}. The GSBIQA approach features a modularized architecture comprising five distinct modules: (1) green saliency detection, (2) saliency-guided data cropping, (3) green BIQA feature extraction, (4) local patch prediction, and (5) saliency-guided global prediction. They are elaborated below.

\subsection{Green Saliency Detection}\label{subsubsec:GreenSaliency}
Drawing inspiration from GreenSaliency~\cite{mei2024greensaliency}, a simplified version is implemented within the proposed green saliency detection model.
Contrary to the original method, which extracts features from five distinct layers or scales, this model selects only the three lowest layers: d4, d8, and d16. Overlapped patches are collected from each of these layers, and two consecutive Saab transforms of sizes $2 \times 2$ and $4 \times 4$ are employed to extract features, resulting in d4, d8, and d16 features derived from the corresponding layers. 
It is important to note that the Saab transforms~\cite{kuo2019interpretable} are applied independently across different layers, unlike the interconnected Saab transform modules found in GreenSaliency.

In the saliency map prediction phase, features from all three layers are concatenated. 
The most significant subset of these features is identified through a Relevant Feature Test (RFT) ~\cite{yang2022supervised}, and subsequently, an XGBoost regressor is utilized to estimate the saliency values for each patch. An up-sampling operation is then executed to generate the predicted saliency map.
The predicted saliency map facilitates data cropping and decision ensemble modules, as detailed in Sections \ref{subsubsec:data_cropping} and \ref{subsubsec:ensemble}, respectively. 
Given the absence of ground truth saliency maps in most IQA datasets, the green image saliency prediction model was pre-trained on the SALICON \cite{jiang2015salicon} dataset, a widely used resource for image saliency prediction studies.

\subsection{Saliency-guided Data Cropping}\label{subsubsec:data_cropping}
In the saliency-guided data cropping process, input images are uniformly cropped into overlapped sub-images of size $256 \times 256$, each retaining the original image's quality score. 
Then, the predicted saliency map is used to calculate an Average Saliency Score (ASS) for each sub-image, defined as:
\begin{equation}
ASS_i = \frac{1}{NM}\sum_{n=1}^{N}\sum_{m=1}^{M}S(n, m),
\end{equation}

where $S(n, m)$ represents the saliency value at pixel coordinates $(n, m)$ in the $i^{th}$ sub-image, and $N \times M$ represents the size of sub-images.
The ASS indicates how well sub-images attract human attention, with higher scores indicating more significant impacts on perceptual judgments. Sub-images with the highest ASS are selected for further analysis, ensuring focus on the most salient parts of the image to optimize subsequent quality assessments.

\subsection{Green BIQA Feature Extraction}\label{subsubsec:BIQA}
In this part, three sets of features are extracted and detailed.

\begin{figure}[!htbp]
\centering
\includegraphics[width=0.7\linewidth]{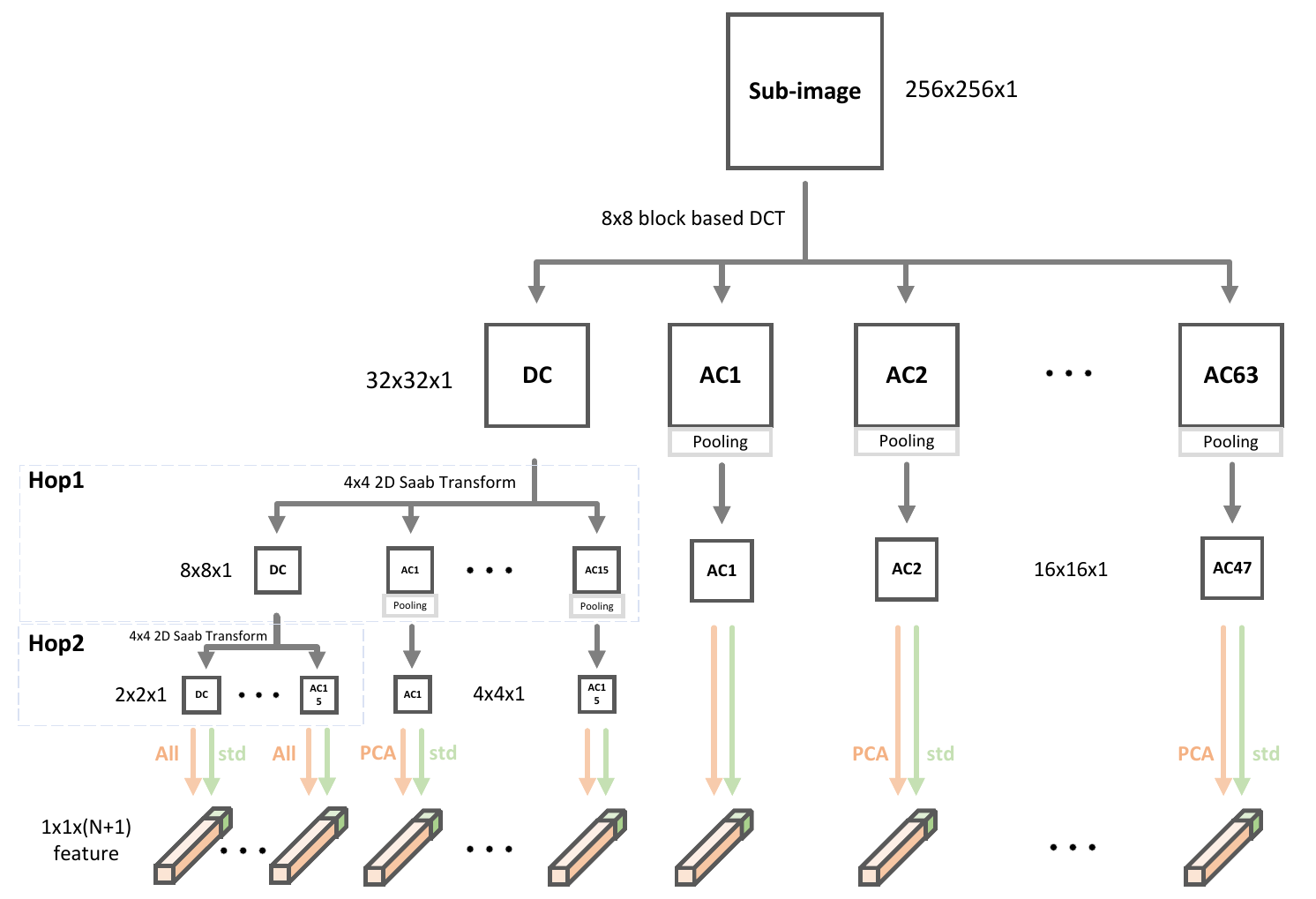}\\
\caption{Structure of spatial features extraction.}
\label{fig:spatial_g}
\end{figure}

\subsubsection{Spatial Features}
The pipeline of spatial feature extraction is depicted in Figure \ref{fig:spatial_g}. 
Initially, input sub-images are segmented into non-overlapping $8 \times 8$ blocks, and Discrete Cosine Transform (DCT) coefficients are computed via a block DCT transform. 
These coefficients are then ordered in a zigzag pattern, producing one DC coefficient and sixty-three AC coefficients labeled AC1 to AC63, then divided into 64 channels. 
For further processing and to decorrelate DC coefficients while deriving higher-level representations, a dual-stage Saab transform approach is utilized, consisting of two successive transformations, referred to as Hop1 and Hop2:
\begin{itemize}
\item Hop1 Processing: The $32 \times 32$ DC coefficients are subdivided into non-overlapping $4 \times 4$ blocks. 
These blocks undergo a Saab transform, resulting in one DC channel and 15 AC channels for each block. 
Each block's $8 \times 8$ DC coefficients are then advanced to Hop2. 
\item Hop2 Processing: A second Saab transform is applied to non-overlapping $4 \times 4$ blocks from DC coefficients from Hop1, yielding one DC and 15 AC channels. 
\end{itemize}
After maximum pooling of Saab coefficients from Hop1 and DCT coefficients from the top layer, we take further steps to reduce the number of representations. This process includes:
\begin{itemize}
\item Computing the standard deviation as shown by the green downward arrow in the diagram.
\item Conduct the PCA transform on spatially adjacent regions of these coefficients as indicated by the orange downward arrow in the diagram. 
\end{itemize}


\subsubsection{Spatio-color Features}
The extraction process for spatial-color features begins by converting sub-images from the YUV color to the RGB color space. 
These converted sub-images form spatial-color cuboids with dimensions 
$H \times W \times C$, where $H$ and $W$ represent the height and width of the sub-image, respectively, and $C=3$ denotes the number of color channels. These cuboids are then input into a two-hop hierarchical structure, similar to that in the spatial features extraction part.
\begin{itemize}
\item In the first hop, the input cuboids are segmented into non-overlapping smaller cuboids of size $4 \times 4 \times 3$. 
A 3D Saab transform is applied to each cuboid independently, resulting in one DC channel and 47 AC channels labeled AC1 through AC47. Each channel maintains a spatial resolution of $64 \times 64$.
\item Given the spatial correlation of the DC coefficients, a 2D Saab transform is utilized in Hop2. 
Here, the DC channel, sized 
$64 \times 64$, is further subdivided into 
$16 \times 16$ non-overlapping blocks, each $4 \times 4$ in dimension. 
\end{itemize}
Similar to those in the spatial features extraction part, all channels of coefficients in each layer are first downsampled by max pooling. Then, PCA coefficients and standard deviation values are computed.

\subsubsection{Saliency Features}
Building on the insights that multi-level features can boost the accuracy of IQA methods, our approach integrates saliency features from multiple layers into the green BIQA feature extraction process. 
These features, initially extracted in the green image saliency detection module and shown as d4 and d8 features in Figure \ref{fig:pipeline}, are resized for compatibility with BIQA dimensions and seamlessly concatenated with BIQA features. 
This integration enhances the feature set by leveraging the spatial coherence and contextual relevance of the saliency data, offering a more comprehensive understanding of image quality and potentially enhancing prediction precision without additional computational overhead.

\subsection{Local Patch Score Prediction}\label{subsubsec:local_predict}
The feature set derived from the green BIQA feature extraction module is characterized by its large dimensionality. 
To address this, we adopt the Relevant Feature Test (RFT), as proposed in \cite{yang2022supervised}, to identify and select the most discriminant features.
RFT is a supervised feature selection technique that evaluates the significance of each feature based on its ability to improve the predictive model's performance. It uses statistical methods to ascertain the relevance of features, ensuring that only the most powerful features are retained for model training.

Due to the lack of Mean Opinion Scores (MOS) for individual sub-images and only a global MOS for the entire image, directly assigning the global MOS to all local sub-images is ineffective. 
To overcome this, we employ an indirect method that allows each sub-image to predict more adaptive local scores while maintaining the average global image score close to the ground truth.
The local quality score of $i^{th}$ sub-image in the $j^{th}$ image and the global quality score of $j^{th}$ image are denoted as $q_{ij}$ and $Q_j$, respectively.
We initiate the training process by uniformly distributing the global quality score across all local sub-images.
An XGBoost regressor is trained to predict $q_{ij}$ for each sub-image. $Q_j$ for the entire image is computed by the average of all $q_{ij}$ from the same image.
The discrepancy between $Q_j$ and its corresponding label is used to update the gradient, which is then propagated back to each local patch.
This iterative process, characterized by successive relaxation and adjustment, converges to a stable set of local quality scores $q_{ij}$.
This approach effectively captures the diverse quality variations across sub-images, providing a more nuanced image quality assessment.

\subsection{Saliency-guided Global Quality Score Prediction}\label{subsubsec:ensemble}
Upon calculating local quality scores for each sub-image, we integrate these scores to form a global quality assessment using an additional XGBoost regressor, which utilizes a detailed set of input features. The feature collection and preparation involve:
\begin{itemize}
\item Collection of all local quality scores from the same image to capture quality variance.
\item Computation of statistical metrics such as average, standard deviation, and maximum value of saliency local scores from the same image.
\item Integration of downsampled d16 features from the green saliency detection module refined by the RFT.
\end{itemize}
The prepared features enable the XGBoost regressor to provide an accurate global image quality prediction, offering a thorough evaluation that reflects both local detail and global perceptual impact.

\section{Experiments}\label{G_sec:experiments}

\subsection{Experimental Setup}\label{G_subsec:setup}

\subsubsection{Datasets}
The green image saliency detection module was pre-trained using the SALICON \cite{jiang2015salicon} dataset.
For the evaluation of the GSBIQA method, we utilized the LIVE-C \cite{ghadiyaram2015massive} and KonIQ-10K \cite{hosu2020koniq} datasets. These datasets are particularly relevant for authentic IQA as they encompass various real-world images captured by users featuring assorted distortions. 

\subsubsection{Evaluation Metrics}

The effectiveness of the proposed method is quantified using two widely recognized metrics: the Pearson Linear Correlation Coefficient (PLCC) and the Spearman Rank Order Correlation Coefficient (SROCC).
PLCC is computed as follows:
\begin{equation}
\textit{PLCC} = \frac{\sum_{i}(p_i - p_m)(\hat{p_i}-\hat{p_m})}
{\sqrt{\sum_{i}(p_i - p_m)^2}\sqrt{\sum_{i}(\hat{p_i}-\hat{p_m})^2}},
\end{equation}
where $p_i$ represents the predicted scores, $\hat{p_i}$ denotes the subjective scores, and $p_m$ and $\hat{p_m}$ are the mean values of the predicted and subjective scores, respectively.
SROCC is calculated as:
\begin{equation}
\textit{SROCC} = 1 -  \frac{6\sum_{i=1}^{L}(m_i - n_i)^2} {L(L^2-1)},
\end{equation}
where $m_i$ and $n_i$ are the ranks of the predicted and actual subjective scores, respectively, and $L$ represents the total number of samples, which corresponds to the number of images evaluated.

\begin{table*}[!htbp]
\centering
\caption{Performance comparison in PLCC and SROCC metrics between our
GSBIQA method and nine benchmarking methods on two IQA databases. The best performance numbers are shown in boldface and "X" denotes the multiple number.}\label{table:BIQA_individual}
\resizebox{\linewidth}{!}{
\begin{tabular}{l  cc cc c c c}
\toprule
\multirow{2}{*}{BIQA Method} & \multicolumn{2}{c}{LIVE-C} & \multicolumn{2}{c }{KonIQ-10K} & \multirow{2}{*}{Model Size (MB)$\downarrow$} & \multirow{2}{*}{GFLOPs$\downarrow$} & \multirow{2}{*}{Inference Time (ms)$\downarrow$}\\
\cmidrule(l){2-3} \cmidrule(l){4-5}
& SROCC$\uparrow$ & PLCC$\uparrow$ & SROCC$\uparrow$ & PLCC$\uparrow$ \\\midrule
BRISQUE~\cite{mittal2012no}  &0.608 &0.629  &0.665 &0.681 &- &- &-\\
CORNIA~\cite{ye2012unsupervised}   &0.632 &0.661 &0.780 &0.795 &7.4 (3.1X) &- &-\\
\hline
BIECON~\cite{kim2016fully}   &0.595 &0.613  &0.618 &0.651 &35.2 (14.6X) &2.82 (2.2X) &26 (1.4X)\\
WaDIQaM~\cite{bosse2017deep}  &0.671 &0.680 &0.797 &0.805 &25.2 (10.5X) &4.38 (3.5X) &26 (1.4X)\\ \hline
PQR~\cite{zeng2018blind}      &0.857 &\textbf{0.882}  &0.880 &0.884 &235.9 (98.3X) &157 (125.6X) &100 (5.3X)\\
DBCNN~\cite{zhang2018blind}     &0.851 &0.869 &0.875 &0.884 &54.6 (22.7X) &242 (193.6X) &55 (2.9X)\\
SGDNet~\cite{yang2019sgdnet} &0.851 &0.872 &0.897 &0.917 &323.6 (134.8X) &260 (208X) &141 (7.4X)\\
HyperIQA~\cite{su2020blindly} &\textbf{0.859} &\textbf{0.882} &0.906 &0.917 &104.7 (43.6X) &145 (116X) &54 (2.8X)\\
TReS~\cite{golestaneh2022no} &0.846 &0.877  &\textbf{0.915} &\textbf{0.928} &582 (242.5X) &290 (232X) &260 (13.7X)\\
\hline
GSBIQA(ours)  &0.830 &0.839 &0.875 &0.883 &\textbf{2.4 (1X)} &\textbf{1.25 (1X)} &\textbf{19 (1X)}\\ 
\bottomrule
\end{tabular}}
\end{table*}

\subsubsection{Implementation Details}
In the training stage, we crop and select 35 sub-images of size $256 \times 256$ for each image in the two authentic datasets.  
The RFT selects 3,000 dimensions of features to do local patch score prediction.
To perform global score prediction, the input features include 35 predicted scores from local patch score prediction, 165 dimensions of features from the statistics of the saliency map, and 600 dimensions of features from saliency features selected by RFT.
We adopt the standard evaluation procedure by splitting each dataset into 80\% for training and 20\% for testing.  
Furthermore, 10\% of training data is used for validation. We ran experiments ten times and reported median PLCC and SROCC values.

\subsection{Experimental Results}
\subsubsection{Benchmarking Methods}\label{subsubsec:benchmarking}
BRISQUE~\cite{mittal2012no} and CORNIA~\cite{ye2012unsupervised} are conventional BIQA methods. BIECON~\cite{kim2016fully} and WaDIQaM~\cite{bosse2017deep} are DL-based BIQA methods without pre-trained models on external datasets.
PQR~\cite{zeng2018blind}, DBCNN~\cite{zhang2018blind},
SGDNet~\cite{yang2019sgdnet},
HyperIQA~\cite{su2020blindly}, and TReS~\cite{golestaneh2022no} are DL-based BIQA methods with pre-trained models on external datasets, adding advanced technologies.

\subsubsection{Comparison among Benchmarking Methods}
The evaluation of two datasets featuring authentic distortions, LIVE-C and KonIQ-10K, reveals that our GSBIQA method surpasses both the conventional BIQA and simple DL-based methods, as detailed in Table \ref{table:BIQA_individual}.
This superior performance underscores the efficacy of the quality-aware features extracted by GSBIQA, highlighting its potential for robust and accurate image quality assessment for real-world image distortions.
While GSBIQA exhibits promising performance, a performance gap remains compared to state-of-the-art DL-based methods. 
However, it is essential to consider the significantly lower model complexity of GSBIQA as a beneficial trade-off. The details of model complexity are discussed in Section \ref{subsec:model_complexity}.

\subsubsection{Qualitative Analysis on Exemplary Images}
\begin{figure}[!htbp]
\centering
\captionsetup{justification=centering}
\includegraphics[width=0.8\linewidth]{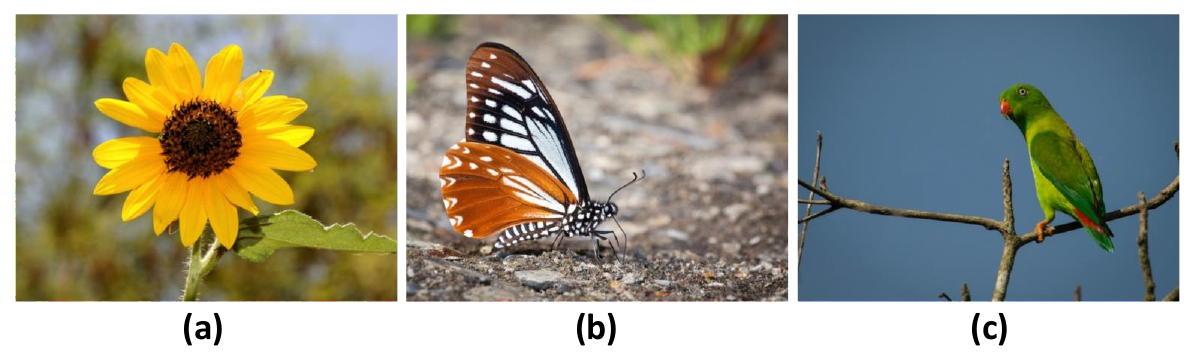}\\
\caption{Successful cases in GSBIQA. \\
(a) MOS(G)=3.80, MOS(P)=3.80, (b) MOS(G)=3.76, MOS(P)=3.76, (c) MOS(G)=3.70, MOS(P)=3.70.
}\label{fig:good_case}
\end{figure}

\begin{figure}[!htbp]
\centering
\captionsetup{justification=centering}
\includegraphics[width=0.8\linewidth]{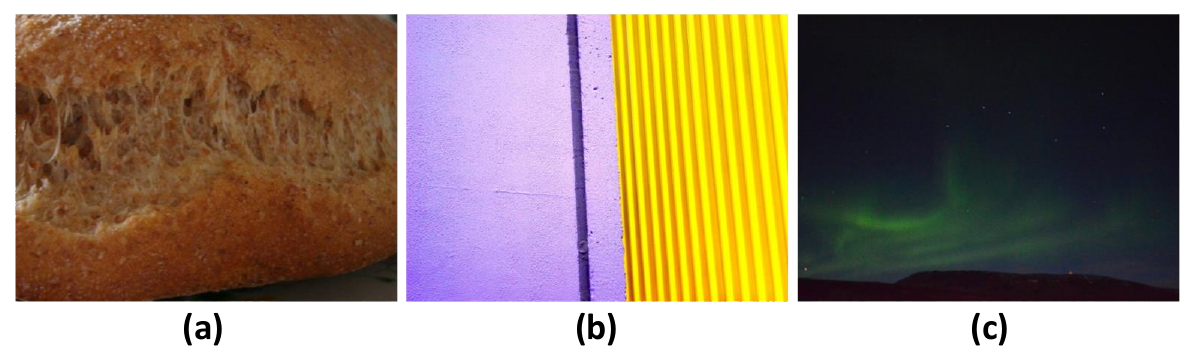}\\
\caption{Failed cases in GSBIQA. \\
(a) MOS(G)=2.61, MOS(P)=2.41, (b) MOS(G)=3.23, MOS(P)=3.00, (c) MOS(G)=2.30, MOS(P)=2.07.
}\label{fig:bad_case1}
\end{figure}

We conducted a qualitative analysis to evaluate the performance of the GSBIQA method by comparing the predicted MOS against the ground truth. 
Figure \ref{fig:good_case} presents exemplary images demonstrating successful outcomes. The ground truth MOS and GSBIQA-predicted MOS values are denoted as MOS(G) and MOS(P), respectively. 
These images are characterized by salient objects, effectively detected by the green image saliency detector. The saliency guidance provided by this component enables GSBIQA to accurately capture these features, significantly enhancing the model's performance. 
These images confirm that GSBIQA is proficient in scenarios where salient objects dominate the visual field.
Conversely, GSBIQA exhibits limitations in images that lack distinct salient objects. Examples of such images include those predominantly featuring full backgrounds or textures, as shown in Figure \ref{fig:bad_case1}. 
In these cases, the absence of transparent, distinguishable objects leads to challenges in saliency detection, resulting in poorer performance.

\subsection{Model Complexity} \label{subsec:model_complexity}
We examine the complexity of BIQA methods from three perspectives: model size, computational complexity as measured by floating-point operations (FLOPs), and inference time, as shown in Table \ref{table:BIQA_individual}.

\subsubsection{Model Size}
Compared to conventional methods such as CORINA, GSBIQA demonstrates superior performance and maintains a smaller model footprint. 
Additionally, GSBIQA outperforms two early-stage DL-based methods, BIECON and WaDIQaM, while achieving a significantly reduced model size. 
In comparison with advanced DL-based methods, namely PQR, DBCNN, SGDNet, HyperIQA, and TReS, GSBIQA offers competitive performance on the LIVE-C and KonIQ-10K datasets, yet with a considerably smaller model size. 
Notably, advanced DL-based methods typically rely on extensive pre-trained networks, often exceeding 100MB, which underscores the efficiency of GSBIQA in maintaining lower resource usage.

\subsubsection{Computational Complexity}
The ``GFLOPs" column in Table \ref{table:BIQA_individual} details the GFLOPs required to process a single image.
Among early-stage DL-based methods such as WaDIQaM and BIECON, although the networks are smaller and require fewer FLOPs, their performance still falls short of GSBIQA. 
Conversely, more advanced DL-based methods like TReSN and HyperIQA, while surpassing GSBIQA in accuracy, also possess considerably larger model sizes and higher computational complexities. 
Specifically, the FLOPs for TReS and HyperIQA are 232 and 116 times greater than those for GSBIQA, respectively.
It is crucial to highlight that non-DL-based methods like GSBIQA might benefit less from GPU acceleration due to a lack of optimized hardware-software integration compared to DL-based methods. 
However, the notably low computational complexity of GSBIQA underscores its potential for effective operation in GPU-supported environments, pending further improvements in third-party libraries and coding optimizations.

\subsubsection{Inference Time}
Inference time is another critical factor, particularly for applications on mobile or edge devices. 
Table \ref{table:BIQA_individual} presents the inference times, measured in milliseconds per image, across various methods on two datasets. 
Each method was evaluated in the same environment using a single CPU.
GSBIQA offers significant advantages over benchmarked methods by jointly considering performance and inference time. 
It is noteworthy that GSBIQA can process approximately 52 images per second using a single CPU, thus meeting the real-time requirements for processing videos at 30 frames per second. 
Furthermore, one can reduce the inference time for GSBIQA through code optimization and the integration of more advanced library support.

\begin{table}[!htbp]
\centering
\caption{Ablation study for GSBIQA on KonIQ-10K dataset.}
\begin{tabular}{c c c  c c}
\toprule
\multicolumn{2}{c}{Saliency Guidance} &\multirow{2}{*}{Local prediction} &\multirow{2}{*}{SROCC} &\multirow{2}{*}{PLCC} \\
\cmidrule(l){1-2} Data Cropping & Global Prediction  \\\midrule
 & & &0.832 &0.845\\\midrule
\tikzcmark & & &0.848 &0.859\\\midrule
 &\tikzcmark & &0.850 &0.860\\\midrule
 & &\tikzcmark &0.862 &0.866\\ \midrule
  \tikzcmark &\tikzcmark & &0.862 &0.868\\\midrule
\tikzcmark &\tikzcmark &\tikzcmark &0.875 &0.883\\
\bottomrule
\end{tabular}
\label{table:ablation}
\end{table}

\subsection{Ablation Study}
We conduct an ablation study in Table \ref{table:ablation} by focusing on three critical components of GSBIQA: data cropping, global quality score prediction, and local patch prediction.
First, the study assessed the impact of each component on its own, as shown in the second to fourth rows.
Both SROCC and PLCC metrics on the KonIQ-10K dataset improved when each component was implemented independently, underscoring the significant role each plays in enhancing model performance.
Next, a combined implementation of data cropping and global quality score prediction, guided by saliency detection, was analyzed. 
The results in the fifth row demonstrated substantial improvements, validating the effectiveness of integrating saliency guidance into these processes.
Finally, we use all components in the sixth row and see that SROCC and PLCC can be further improved to reach the highest value.

\section{Conclusion and Future Work}\label{G_sec:conclusion}
A novel and lightweight saliency-guided BIQA method,
called GSBIQA, was proposed. GSBIQA outperforms conventional and early-stage DL-based methods on two authentic distortion datasets. It also offers competitive performance with a significantly smaller model size than state-of-the-art methods. It processes 52 images per second using only CPU resources, making it well-suited for mobile and edge devices.
Future research directions include enhancing the adaptability of the saliency guidance system to improve performance across varied image types and expanding GSBIQA for blind video quality assessment (BVQA).

\bibliographystyle{unsrt}  
\bibliography{references}

\end{document}